\documentclass[twocolumn, trackchanges]{aastex631}
\usepackage{amsmath}
\usepackage{subfigure}
\usepackage{ulem}
\usepackage{wasysym}
\usepackage{comment}
\usepackage{gensymb}
\usepackage{url}

\shortauthors{Louden, Laughlin, and Millholland}
\shorttitle{Dissipation Regimes in Exoplanets}

\begin{document}

\title{Tidal Dissipation Regimes Among the Short-Period Exoplanets}

\correspondingauthor{Emma M. Louden}
\email{emma.louden@yale.edu}

\author[0000-0003-3179-5320]{Emma M. Louden}
\affiliation{Yale University, 52 Hillhouse Avenue, New Haven, CT 06511, USA}

\author[0000-0002-3253-2621]{Gregory P. Laughlin}
\affiliation{Yale University, 52 Hillhouse Avenue, New Haven, CT 06511, USA}

\author[0000-0003-3130-2282]{Sarah C. Millholland}
\affiliation{MIT Kavli Institute for Astrophysics and Space Research, Massachusetts Institute of Technology, Cambridge, MA 02139, USA}

\begin{abstract}

The efficiency of tidal dissipation provides a zeroth-order link to a planet's physical properties. For super-Earth and sub-Neptune planets in the range $R_{\oplus}\lesssim R_p \lesssim 4 R_{\oplus}$, particularly efficient dissipation (i.e., low tidal quality factors) may signify terrestrial-like planets capable of maintaining rigid crustal features. Here we explore global constraints on planetary tidal quality factors using a population of planets in multiple-planet systems whose orbital and physical properties indicate susceptibility to capture into secular spin-orbit resonances. Planets participating in secular spin-orbit resonance can maintain large axial tilts and significantly enhanced heating from obliquity tides. When obliquity tides are sufficiently strong, planets in low-order mean-motion resonances can experience resonant repulsion (period ratio increase). The observed distribution of period ratios among transiting planet pairs may thus depend non-trivially on the underlying planetary structures. We model the action of resonant repulsion and demonstrate that the observed distribution of period ratios near the 2:1 and 3:2 commensurabilties implies $Q$ values spanning from $Q\approx 10^1-10^7$ and peaking at $Q \approx 10^6$. This range includes the expected range in which super-Earth and sub-Neptune planets dissipate ($Q \approx 10^3 - 10^4$). This work serves as a proof of concept for a method of assessing the presence of two dissipation regimes, and we estimate the number of additional multi-transiting planetary systems needed to place any bimodality in the distribution on a strong statistical footing.

\end{abstract}


\section{Introduction} \label{sec:intro}

The growing census of short-period exoplanets with joint mass and radius measurements hints at two distinct populations: super-Earths that may resemble scaled-up terrestrial worlds, and sub-Neptunes, characterized by structurally significant H/He gas envelopes. The presence of the distinct populations follows naturally from the bimodal radius distribution among the \textit{Kepler}-detected \citep{2010Sci...327..977B} planets \citep{2017AJ....154..109F, 2018MNRAS.479.4786V, 2018AJ....156..264F} which exhibit a well-established paucity in the range 1.5$R_\oplus \lesssim R_{\rm P} \lesssim 2 R_\oplus$. 

The gap in the planet radius distribution was predicted by models that draw on the photoevaporation of planetary atmospheres \citep{2012ApJ...761...59L, 2013ApJ...775..105O, 2014ApJ...795...65J}.  Extreme ultraviolet and X-ray radiation from the host star heats the upper regions of the planetary atmosphere and drives outflows \citep{2009ApJ...693...23M, 2013ApJ...775..105O,2013ApJ...776....2L}. Another energetic driver that may spur mass loss is the heat of formation of the solid core \citep{2018MNRAS.476..759G, 2019MNRAS.487...24G}. Because thermally-driven mass loss is likely a threshold process, the smaller-radius peak in the radius distribution may comprise rocky planets whose atmospheres have been entirely stripped. In contrast, the planets in the larger peak have retained significant atmospheres throughout the parent star's early high-activity phase and beyond. In many cases, the energy input inflates the retained envelopes, enhancing contribution to the larger-radius peak in the distribution \citep{2023MNRAS.518.1683K}. Another hypothesis is that the sub-Neptune type planets are, in fact, water worlds rather than H/He-rich \citep{2019PNAS..116.9723Z, 2022Sci...377.1211L}.

If the planets on either side of the radius gap have fundamentally different structures, it is reasonable to expect a corresponding bimodal distribution of tidal quality factors. This paper assesses whether evidence of a distinct split in planetary composition can be discerned in the current catalog of transiting extrasolar planets. 

Generically, the tidal quality factor, $Q$, measures the efficiency of tidal dissipation with
\begin{equation}
     Q^{-1} = -\frac{1}{2\pi E_0}\oint \frac{dE}{dt}\,dt\, ,
\end{equation}
where $E_0$ is the maximum energy stored in the tidal distortion of the planet and the integral represents the energy lost in one cycle \citep{1966Icar....5..375G}. Terrestrial bodies in the Solar System tend to have $Q_{\rm terrestrial}\sim 10^2$, while Uranus and Neptune are at least two orders of magnitude less dissipative ($Q_{\rm ice-giant}\gtrsim 10^4$) \citep{1966Icar....5..375G}. With the exception of Earth and the Moon, however, planet-wide energy dissipation rates are difficult to measure with high accuracy. For the Moon, a precise value $Q_{\leftmoon}\sim 38$ was inferred via laser ranging \citep{2015JGRE..120..689W}, whereas for Earth, the handbook value is $Q_{\oplus}=12$  \citep[e.g.,][]{Yoder1995}. A recent estimate reported by \cite{2016CeMDA.126..145L} draws on astrometric measurements to obtain $Q\sim100$ for Mars. Numerical simulations that reconstruct the past dissipative orbital evolution of the Uranian and Neptunian satellites suggest a tidal $Q$ between 15,000 and 20,000 for Uranus \citep{2020PSJ.....1...22C} and a $Q$ between 9,000 and 36,000 for Neptune \citep{2008Icar..193..267Z}. 

Currently, there are limited constraints on tidal $Q$ values for the lower-mass ($M_{\rm P}<30\,M_{\oplus}$) exoplanets. Planets in this category that are individually constrained include GJ 876 d  \citep{2018AJ....155..157P}, and GJ 436 b \citep{2017AJ....153...86M}. The former was measured by considering the competing roles of eccentricity damping by tides and eccentricity excitation by the Laplace-like resonance chain in the system, whereas the latter was constrained by fitting the planet's emission spectrum to detailed atmospheric models that include the outward flux of interior heat generated by tidal dissipation within the planet.

The efficiency of tidal dissipation depends on the body's rheological properties \citep{2015JGRE..120..689W} whose global expressions are reflected in the planetary Love number, rigidity, bulk modulus, and overall density. In particular, $Q$ is anti-correlated with the core radius and shear modulus \citep{2016CeMDA.126..145L}, implying that terrestrial bodies dissipate more efficiently.

\begin{figure}
    \centering
    \includegraphics[width=0.5\textwidth]{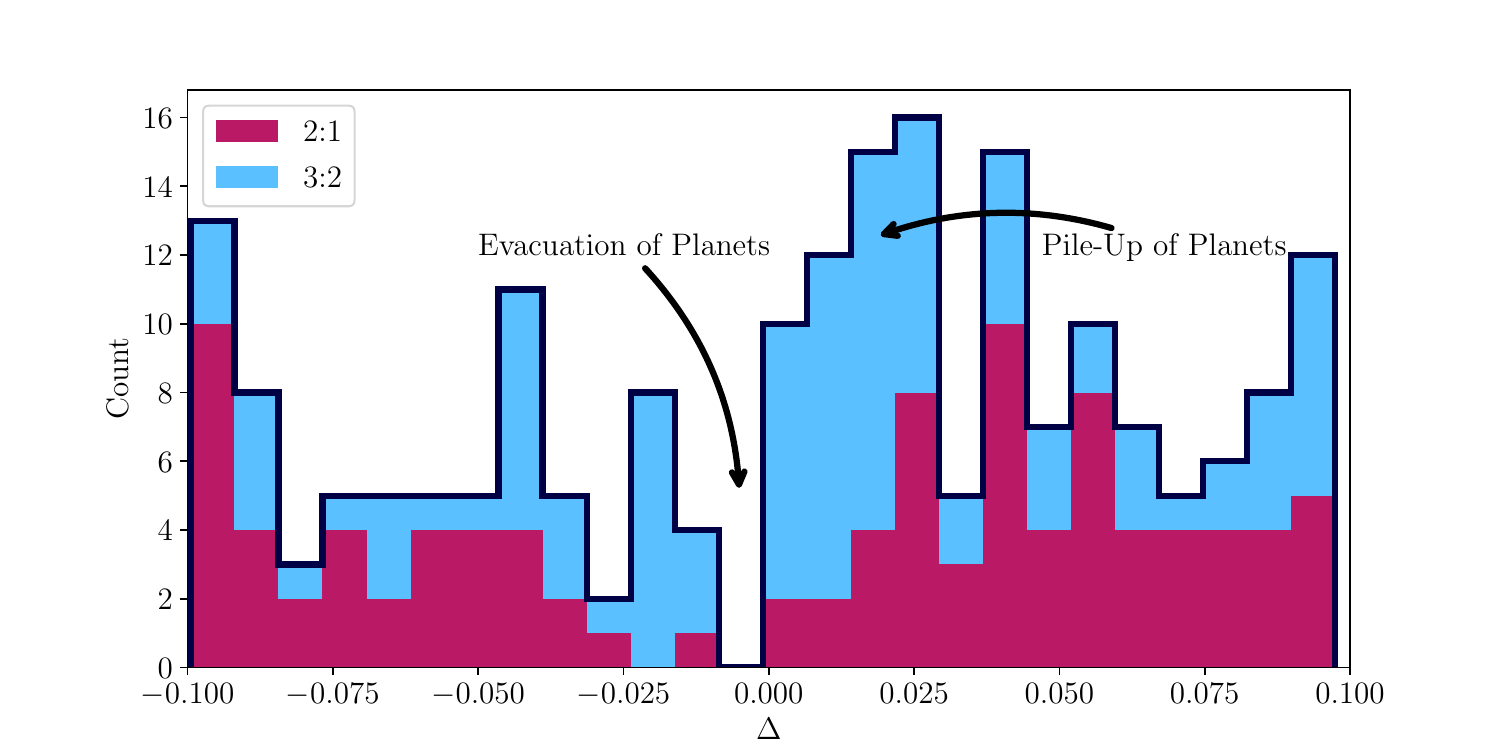}
    \caption{\textbf{Fractional distance from resonance for planet pairs near first-order mean motion resonances.} Multi-transiting exoplanetary systems near the 3:2 and 2:1 commensurabilities tend to accumulate just wide of the exact resonances (which occur for $\Delta = 0$). There is both an over-density at ratios $\Delta > 0$  and a paucity for $\Delta < 0$.}
    \label{fig:period_ratio_hist}
\end{figure}

Given that some short-period exoplanets may be subject to strong tidal forcing, the impacts of their dissipative properties may be evident within demographic features of the population such as the distribution of orbital period ratios between pairs of planets within the same system. The period ratio distribution observed by NASA's \textit{Kepler} Mission revealed that most ($\sim80-90\%$) planets in compact multiple-planet systems have period ratios that are inconsistent with participation in low-order mean-motion orbital resonance. There is, however, a notable excess of planet pairs with period ratios just above the 2:1 and 3:2 commensurabilities and a dearth just below \citep{2011ApJS..197....8L, 2014ApJ...790..146F}. 

Figure \ref{fig:period_ratio_hist} illustrates these features in the period ratio distribution. The pile-up and evacuation can be seen distinctly when the period pairs are represented as a function of $\Delta$, the fractional distance of a  pair of periods $P_2 > P_1$ from exact $(j+1):j$ commensurability
\begin{equation}
\Delta_{j+1:j} = \frac{j}{j+1}\frac{P_2}{P_1} - 1.
\end{equation}

\begin{figure}
    \centering
    \includegraphics[width=0.5\textwidth]{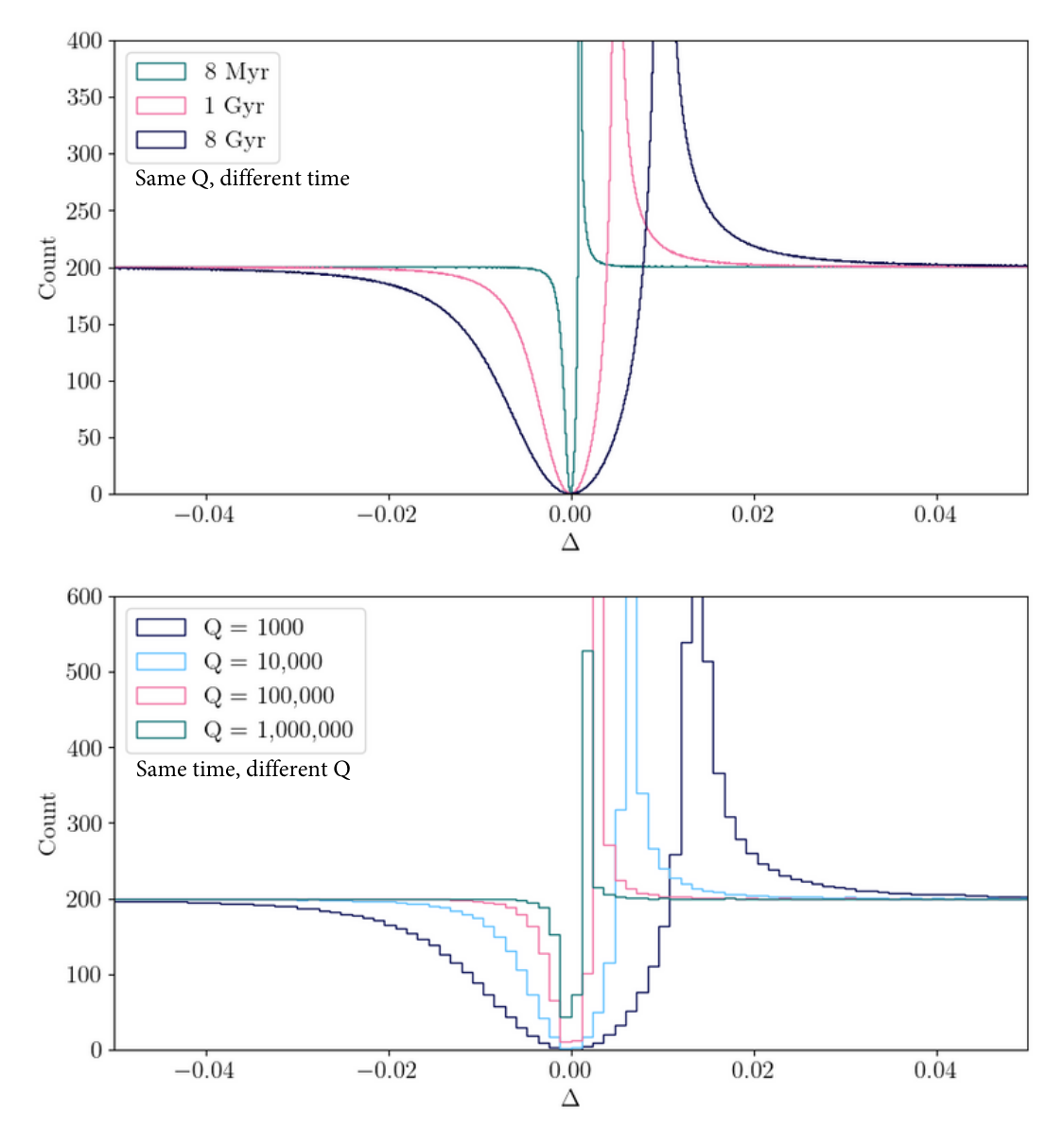}
    \caption{\textbf{Simulated evolution of period ratios for a synthetic planet population.} \textit{Top:} Time evolution of a population of planets initially uniformly distributed in $\Delta$ due to the action of the model encapsulated by Equation \ref{eq:naive_dmig} from \cite{2012ApJ...756L..11L} which assumes a fiducial $Q$ of 10. \textit{Bottom:} The time evolution over 2 billion years of four populations parameterized by $Q$, and spaced initially uniformly in $\Delta$. Systems with lower $Q$ are more dissipative than those with higher $Q$ and evacuate the region near $\Delta = 0$ more rapidly.}
    \label{fig:combined}
\end{figure}

\cite{2012ApJ...756L..11L} \citep[see also][]{2013AJ....145....1B} showed that when two planets in mean-motion resonance harbor both a significant angular momentum deficit and significant dissipation, the planets will migrate apart (or repel each other) in period ratio space with the evolution exhibiting an overall time dependence
\begin{equation}
    \label{eq:naive_dmig}
    \Delta_{mig} \propto (t/\textrm{Gyr})^{1/3}\, ,
\end{equation} 
where the constant of proportionality depends both on the orbits and the interior planetary properties for a given planetary pair.
In \S 4, we examine this evolution of the $\Delta$-value for a planet pair in more detail, extending it to account for both eccentricity and obliquity tides. Here, we note that when applied to a population of planets with an initially uniform $\Delta$ distribution, this heuristic can generate the features exhibited by the \textit{Kepler} planets. For each individual resonant pair in the initial distribution, orbital energy is lost to heat, the two planets repel each other, the orbits circularize, and the period ratio increases. Figure \ref{fig:combined} shows this process acting over time on a population. 

Although resonant repulsion can qualitatively explain the period ratio features, the source of dissipation has been difficult to pin down. Eccentricity tides alone are insufficient to produce the resonant features because the dissipation associated with the typical eccentricities within compact multi-planet systems is not strong enough to increase the period ratios by the observed amount \citep[e.g.][]{2015MNRAS.453.4089S}. Possible solutions include dissipation from the disk \citep{2020MNRAS.495.4192C, 2022arXiv221115701C} or dissipation from obliquity tides, as proposed by \cite{2019NatAs...3..424M} and explored in detail in this work. 

Obliquity tides result when the spin axis and the orbital angular momentum vector of a planet are misaligned. Dissipation occurs as the planet's tidal bulge is continually forced across the planet during the course of an orbit \citep[e.g.][]{2005ApJ...628L.159W}. Obliquity tides are an attractive mechanism to explain the near-resonant features in the period ratio distribution. First, as shown by \cite{2019NatAs...3..424M}, close-in multiple-planet systems are susceptible to ``secular spin-orbit resonances,'' which involve a commensurability between a planet's spin-axis precession rate and its orbital precession rate (see Section \ref{sec:CassiniObli} for a detailed explanation of this resonance). Such resonances allow obliquity tides to operate over extended periods, thereby providing a potential avenue for generating sufficient dissipation for planet pairs to experience resonant repulsion. Second, the degree of extra dissipation offered by obliquity tides is sufficient to match the observed offsets between the period ratio accumulations and the exact period ratio commensurabilities (discussed more in Section \ref{sec:CassiniObli} and in Figure \ref{fig:obliquity heatmap}). 

In this paper, we study how the period ratio distribution evolves when a significant number of its constituent planet pairs are participating in orbital mean-motion resonance while simultaneously having one or both pair members trapped in secular spin-orbit resonance. In all likelihood, a combination of disk effects and tides could be responsible for sculpting the observed period ratio features. However, given the complication of modeling both disk-driven and tidally-driven effects, our goal is to isolate the effects of tides here and see what inferences can or cannot be made about the $Q$ distribution. \cite{2019NatAs...3..424M} explored whether an analysis of this type could be used to glean assessments of the range of $Q$-values and internal planetary structures that typify the \textit{Kepler} planets. That work performed a cursory analysis, however, and it did not incorporate the growing sample of transiting multiple-planet systems emerging from the NASA TESS Mission \citep{2015JATIS...1a4003R}. It is thus helpful to revisit the issue in more detail. \textit{Specifically, do we see evidence of statistically distinct dissipation regimes among the population of short-period exoplanets?}

The plan of this paper is as follows. In Section \ref{sec:CassiniObli}, we discuss the related physics and dynamics of obliquity tides. Section \ref{sec:planetsample} describes the planet sample used in our analysis. In Section \ref{sec:exoQ}, we provide calculations of the tidal $Q$ distribution of the sample and constrain the false alarm probability of our signal. Section \ref{sec:discussion} reviews the conclusions and discusses the findings in the context of future observations.

\section{Tidal and Spin Dynamical Models}\label{sec:CassiniObli}

\begin{figure}
    \centering
    \includegraphics[width=0.45\textwidth]{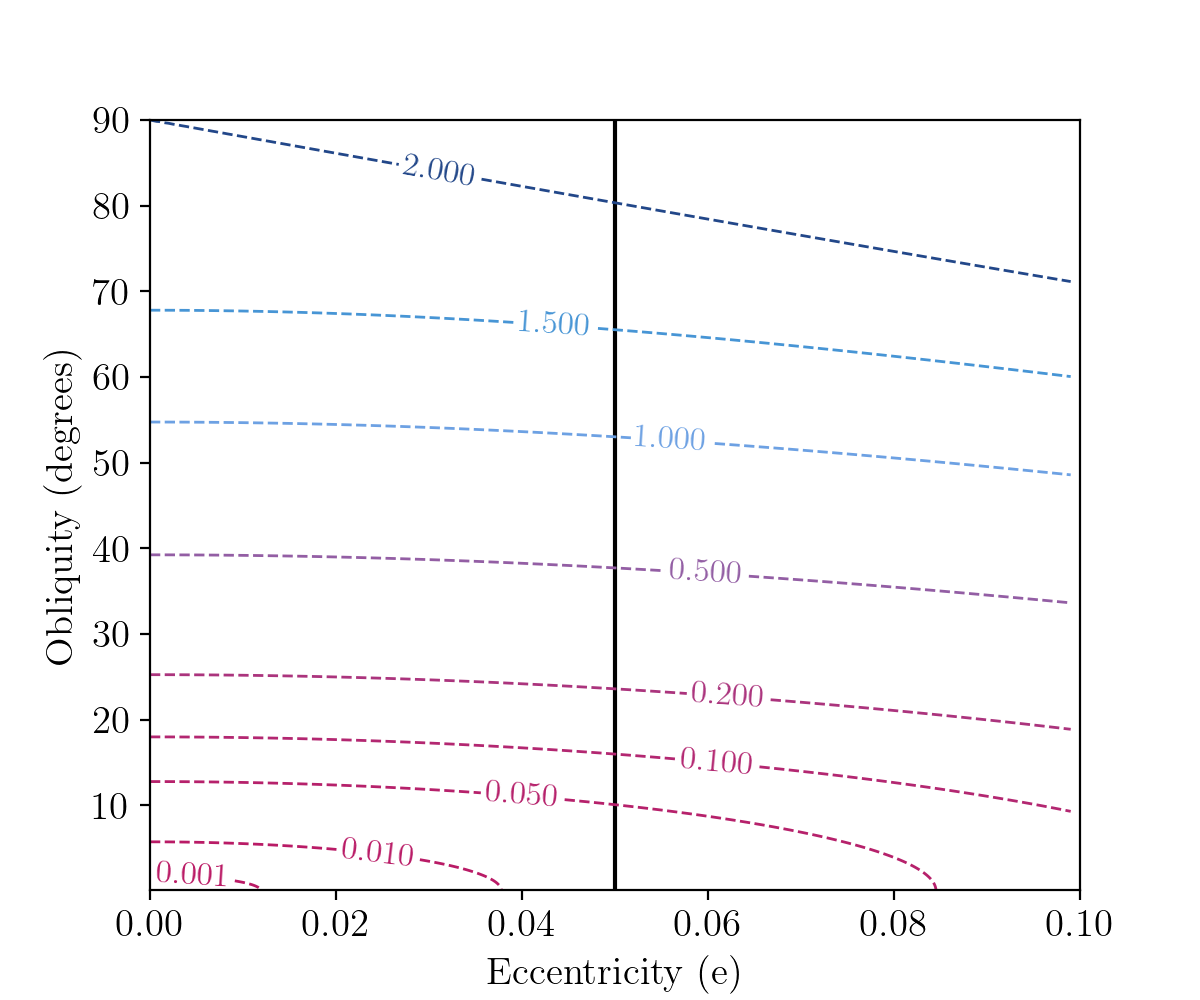}
    \caption{\textbf{Dependence of normalized tidal dissipation rate on eccentricity and obliquity.} The map shows contours of the normalized tidal dissipation rate ($\dot{E}(e,\epsilon)/K_p$ from equation  \ref{eq:Edot}). Dissipation is enhanced by several orders of magnitude for a given eccentricity as obliquity increases. Marked on the plot is the scale parameter, $\sigma=0.05$, that provides the best-fit Rayleigh distribution modeling the eccentricity distribution of compact multi-transiting multi-planet systems \citep{Mills_2019}.}
    \label{fig:obliquity heatmap}
\end{figure}

We employ a simple viscoelastic model for the response of a fluid planet to gravitational perturbations. \cite{1981A&A....99..126H} holds that deformations from equilibrium are characterized by a constant time lag between the radial line to the perturber and the long axis of the tidal bulge. With this simplifying approximation (which holds to second order in the orbital eccentricity), and with eccentricity $e$, mean motion, $n$,  and obliquity, $\epsilon$, the tidal luminosity of the planet is given by \citep{Leconte_2010}

\begin{equation}
\dot{E}(e,\epsilon)=
K_p\frac{2}{1+\cos^2\epsilon}
[\sin^2{\epsilon}+e^2(7+16\sin^2\epsilon)]\, ,
\label{eq:Edot}
\end{equation}
where
\begin{equation}
K_p=\frac{3n}{2}\frac{k_2}{Q}
\left(\frac{GM_{\star}^2}{R_p}\right)
\left(\frac{R_p}{a}\right)^6\, ,
\end{equation}
and where $k_2$ is the Love number, with small $k_2$ implying significant central concentration \citep{1966AJ.....71..891C}. Equation \ref{eq:Edot} indicates that if a planet is in a high-obliquity state, then significant dissipation can be present even if the orbit is circular (see Figure \ref{fig:obliquity heatmap}). In general, tidal dissipation damps the obliquity and realigns the spin axis. However, if a dynamical mechanism exists to maintain the planet's obliquity in the face of damping, then the planet can continuously tap orbital energy to generate tidal luminosity.

Higher obliquities significantly boost the tidal luminosity. \cite{2012ApJ...756L..11L} found that tidal quality factors of order $Q=10$ are required for resonant repulsion to explain the period ratio distribution if eccentricity tides provide the dissipation. If one or both planets maintain high obliquity, however, dissipation for given $Q$ can increase substantially. For instance, at $e = 0.01$, dissipation increases by a factor of 100 for $\epsilon=0\degree\rightarrow 15\degree$ and by 1000 for $\epsilon=0\degree \rightarrow 45\degree$.

\cite{2019NatAs...3..424M} noticed that many planets in the population of compact multi-planet systems have orbital and physical properties that render them susceptible to capture into a secular spin-orbit resonance \citep{1966AJ.....71..891C}, which can excite and maintain high obliquities. In such a configuration, the average precession rate of the planet's spin axis is driven to match the nodal regression rate (or a Fourier component thereof) of the planet's orbital plane. The damped fixed point is known as a Cassini State \citep{Peale1969}, and any adiabatic evolution in the nodal regression rate, $g = \dot{\Omega} < 0$, forces a corresponding evolution in the planetary obliquity, $\epsilon$, such that the relation 

\begin{equation}
\label{eq: Cassini state criterion}
    g\sin(\epsilon - I) + \alpha \cos{\epsilon}\sin{\epsilon} = 0,
\end{equation}

is maintained. Here, $I$ is the planet's orbital inclination and $\alpha$ is the spin-axis precession constant, given by

\begin{equation}
\alpha =\frac{1}{2}\frac{M_\star}{M_p}\left(\frac{R_p}{a}\right)^3\frac{k_2}{C}\omega_p\, ,
\end{equation}

with $\omega_p$ the planet's angular spin frequency and $C$ the dimensionless moment of inertia factor (e.g. $C=2/5$ for a uniform density sphere). In the limit of small inclinations and large obliquities, $\epsilon \gg I$, the Cassini state criterion reduces to
\begin{equation} \label{eq:g_simplified}
|g| \approx \alpha \cos \epsilon \, .
\end{equation}

If $|g| /\alpha$ evolves adiabatically and approaches unity from above, capture into secular spin-orbit resonance occurs, and continued decrease of $|g| /\alpha$ increases $\epsilon$, the planetary obliquity. 

When pair members' obliquities are sufficiently large, planet pairs in low-order mean-motion resonance experience tangible period ratio increases as orbital energy is lost to heat. The rate of evolution depends inversely on $Q$, offering a potential opportunity to probe planetary tidal dissipation regimes. Our goal is to explore whether the currently known population of compact multi-transiting multi-planet systems \textit{already} allows this probe to be made.

\section{Planet Sample} \label{sec:planetsample}
The current aggregate of known transiting exoplanetary systems are listed on the NASA Exoplanet Archive (NEA). We select systems with planetary radii, $R_{p}$, having $R_{p}<9.14 \, R_{\oplus}$ (equivalent to Saturn's radius) and orbital periods, $P< 300\,$d. Additionally, we required the systems to have more than one transiting planet, not be flagged as controversial, and have reported values for semi-major axis, planetary radii, and planetary masses. We used the NEA-reported planetary masses (or $M$, or $M\sin{i}$ in that order of preference, depending on availability; no masses came from mass-radius relationships), stellar ages, and periods. If no value was reported for the age, we adopted a randomly selected age from the cleaned \cite{2020AJ....160..108B} planet sample used in \cite{2021ApJ...920L..34M}. If no value was reported for eccentricity, we adopted a randomly selected eccentricity from a Rayleigh distribution with scale factor $\sigma=0.05$ \citep{Mills_2019}. The NEA catalog is derived from heterogeneous sources, and in many cases, the reported values are subject to considerable uncertainty. Our analysis is thus an initial exploration and subject to refinement as the accuracy and quantity of measurements in the catalog increase.

\begin{figure}
    \centering
    \includegraphics[width=0.45\textwidth]{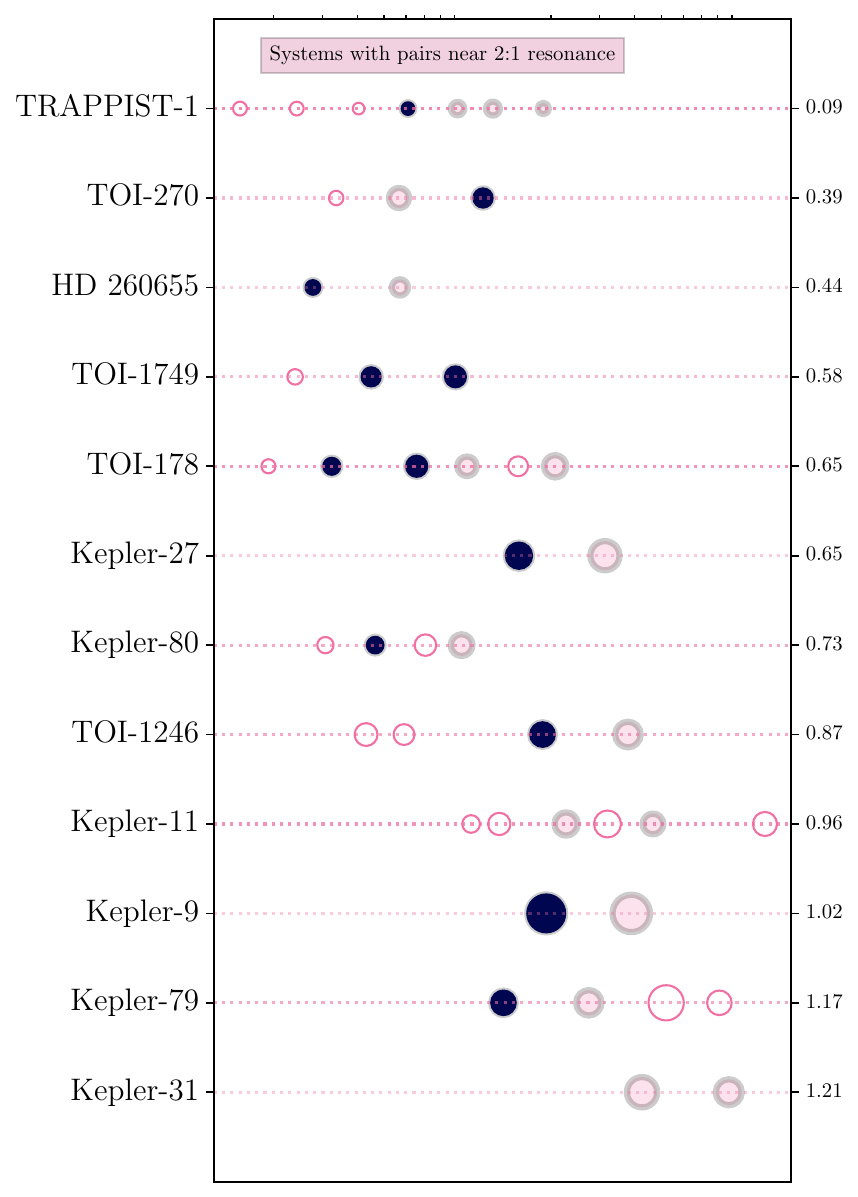}
    \caption{\textbf{Systems with multiple transiting planets displaying period ratios near 2:1.} Symbol sizes are proportional to reported planetary radii, and the systems are ordered by increasing stellar mass (printed on the right axis). The navy-shaded planets are both near 2:1 MMR and have inferred dissipation factors $3<\log_{10}{Q}<6$. The light-pink-shaded points with gray borders represent planets near the 2:1 commensurability but with a $\log_{10}{Q}$ outside of $3-6$.}
    \label{fig:Q_data21}
\end{figure}

\begin{figure}
    \centering
    \includegraphics[width=0.45\textwidth]{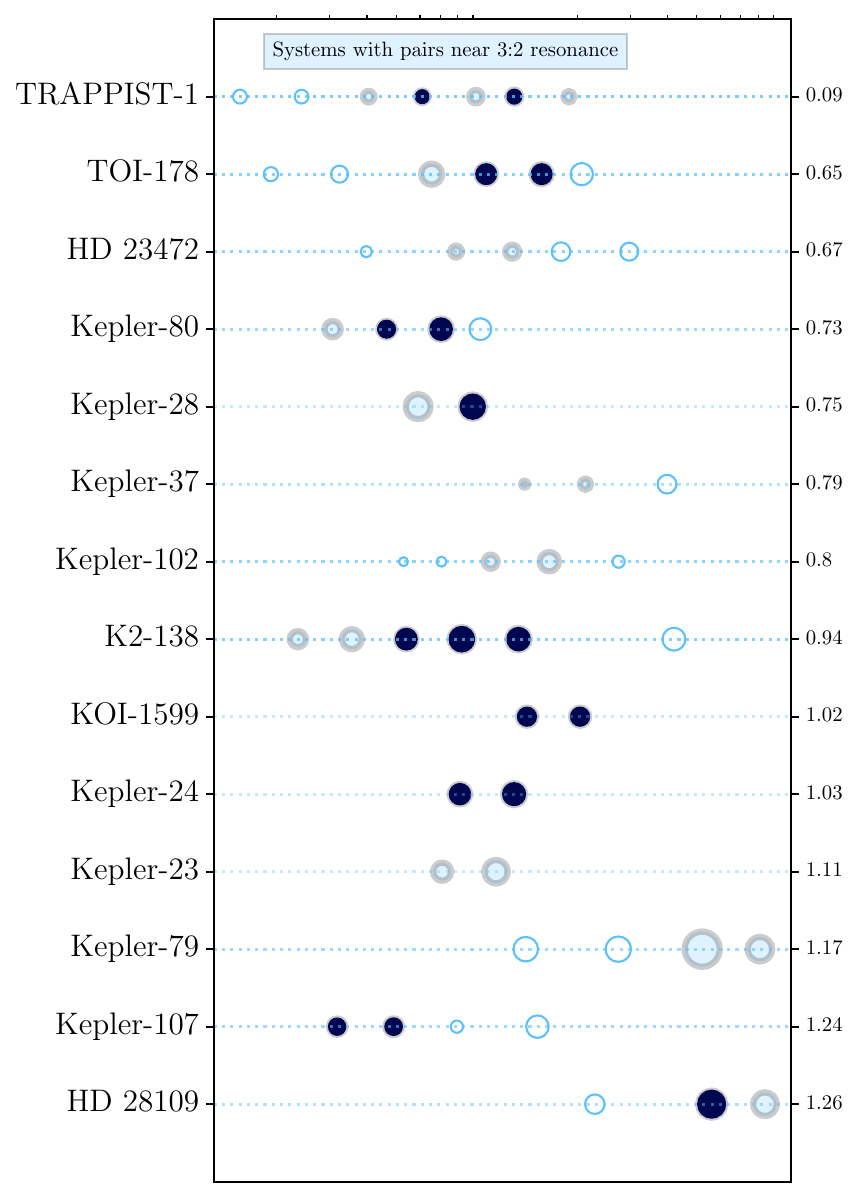}
    \caption{\textbf{Systems with multiple transiting planets displaying period ratios near 3:2.} The features are exactly analogous to Figure \ref{fig:Q_data21}, except for 3:2.} 
    \label{fig:Q_data32}

\end{figure}

In total, 193 planets in 90 systems satisfy the constraints. The planets were further filtered by proximity to the 2:1 and 3:2 orbital commensurabilities. If an outer planet has a period ratio with an interior planet in the range $2.0<\delta<2.1$ or $1.5<\delta<1.6$, it is marked as being sufficiently near a resonant commensurability. A total of $N=35$ planet pairs contribute to the filtered sample. Four systems have planet pairs near both the 3:2 and 2:1 commensurablities. The contributing systems are shown in Figures \ref{fig:Q_data21} and \ref{fig:Q_data32}.

\section{Tidal Dissipation Regimes for Exoplanets} 
\label{sec:exoQ}
To estimate the $Q$ distribution for the planets in our sample, we relate $Q$ to $\Delta$, a planet pair's fractional distance from perfect commensurability.  To achieve this, we adopt a combination of the resonant repulsion formulation in \cite{2012ApJ...756L..11L} \citep[described independently by][]{2013AJ....145....1B} and the orbital evolution described by equilibrium tides \citep{1981A&A....99..126H} extended to arbitrary obliquities by \citet{Leconte_2010}. With this framework, the time evolution of a resonant planetary pair under the action of dissipation can be approximated as

\begin{equation} \label{eq:mig}
    \Delta_{mig} = \left(\frac{9}{4} \mu_1^2 \Gamma t\right)^{1/3}\, ,
\end{equation}

where $\mu_j=m_j/M_*$, and 
\begin{equation}
    \Gamma = (2+\beta)(\gamma_{e1}f_1^2\beta + \gamma_{e2}f_2^2) + \frac{\gamma_{a1}f_1^2\beta^2-\gamma_{a2}f_2^2}{2}.
\end{equation}
where $\beta \equiv \frac{\mu_2 \sqrt{a_2}}{\mu_1 \sqrt{a_1}}$, and $\gamma_e$, $\gamma_a$ are defined below. The terms $f_1$ and $f_2$ are Laplace coefficients (see \citealt{Murray1998} and \citealt{Deck_2013}).

In order to incorporate obliquity tides, the definitions for $\gamma_a$ and $\gamma_e$ need to be generalized. Using \cite{Leconte_2010}'s formulation, we have,

\begin{equation}\label{eq:leconte1}
\frac{1}{\gamma_e} = \frac{1}{e}\frac{de}{dt} = \frac{11a}{G M_{*}M_{p}} K_p \left(\Omega_e(e)x_p\frac{\omega_p}{n}-\frac{18}{11} N_e(e)\right) \, ,
\end{equation}
and
\begin{equation}\label{eq:leconte2}
\frac{1}{\gamma_a} = \frac{1}{a}\frac{da}{dt} =  \frac{4a}{G M_{*}M_{p}} K_p \left(N(e)x_p\frac{\omega_p}{n}-N_a(e)\right)  \, ,
\end{equation}
where $x_p = \cos{\epsilon}$, and $\Omega_e(e), N_e(e), N(e),$ and $N_a(e)$ are eccentricity functions (below) \citep[as derived, for example, by][]{1981A&A....99..126H}.

\begin{equation}
    N(e) = \frac{1+ \frac{15}{2}e^2 + \frac{45}{8}e^4 + \frac{5}{16}e^6}{(1-e^2)^{6}}
\end{equation}
\begin{equation}
    N_a(e) = \frac{1+ \frac{31}{2}e^2 + \frac{255}{8}e^4 + \frac{185}{16}e^6 + \frac{25}{64}e^8}{(1-e^2)^{15/2}}
\end{equation}
 \begin{equation}
    \Omega_e(e) = \frac{1+ \frac{3}{2}e^2 + \frac{1}{8}e^4}{(1-e^2)^{5}}
\end{equation}
\begin{equation}
    N_e(e) = \frac{1+ \frac{15}{4}e^2 + \frac{15}{8}e^4 + \frac{5}{64}e^6 }{(1-e^2)^{13/2}}
\end{equation}
Next, define
\begin{equation}
K_p = \frac{X}{Q}, \gamma_e = \frac{C_e X}{Q}, \gamma_a = \frac{C_a X}{Q}
\end{equation}
 where $C_a$, $C_e$ and $X$ incorporate the system parameters and the functions of eccentricity and obliquity. In addition to specifying the parameters discussed in Section \ref{sec:planetsample}, we need to adopt values for $\omega_p$ and $x_p$.

First, we look at $\omega_p$. Spin synchronization occurs rapidly in comparison to the tidal evolution timescale, so we assume pseudo-synchronous rotation,
 \begin{equation}
\omega_p = \frac{N(e)}{\Omega_e(e)}\frac{2\cos{\epsilon}}
{1+\cos^2{\epsilon}}n \, .
\label{eq:pseudosynch}
\end{equation}
 For $x_p$, we note that in the limit of small $e$, Equation \ref{eq:Edot}
reduces to  
\begin{equation}
\dot{E}(e,\epsilon)=
K\frac{2\sin^2{\epsilon}}{1+\cos^2\epsilon}.
\label{eq:Edot_ecc0}
\end{equation}
As illustrated in Figure \ref{fig:obliquity heatmap}, when $\epsilon > 30\degree$, the value of $\dot{E}/K$ varies by no more than a factor of two. We thus adopt the simplifying assumption that $\epsilon = 60\degree$ for planets for which obliquity tides are the primary source of dissipation.

If we assume that one of the two planets in a pair is trapped in secular spin-orbit resonance and is responsible for the dissipative evolution of the period ratio, then we can obtain an estimate for that planet's individual $Q$. For the pairs of planets that we observe, however, it is unknown which planet(s) (if either) is responsible for obliquity-driven repulsion that may have occurred. We, therefore, consider the options independently and, afterwards, average the results.

\subsection{Estimating Q}\label{sec:inner}
If the inner planet alone is caught in a high-obliquity Cassini State, then we have
$\gamma_{e2} = \gamma_{a2} = 0$, and the expression for $\Gamma$ becomes 
\begin{equation}
    \Gamma = (2+\beta)\left(\frac{C_{e1} X_1}{Q_{1}}f_1^2\beta\right) + \frac{1}{2} \frac{C_{a1} X_1}{Q_{1}}f_1^2\beta^2
\end{equation}
Substituting for $\Gamma$ and rearranging gives an estimate for the inner planet's tidal quality factor
\begin{equation}
   Q_1   = \frac{9\mu_1 t}{4} \Delta_{mig}^{-3}\left[(2+\beta)\left(C_{e1} X_1 f_1^2\beta\right) + \frac{1}{2} C_{a1} X_1 f_1^2\beta^2\right]
   \label{eq:innerQ}
\end{equation}

\begin{figure*}
    \centering
\includegraphics[width=\textwidth]{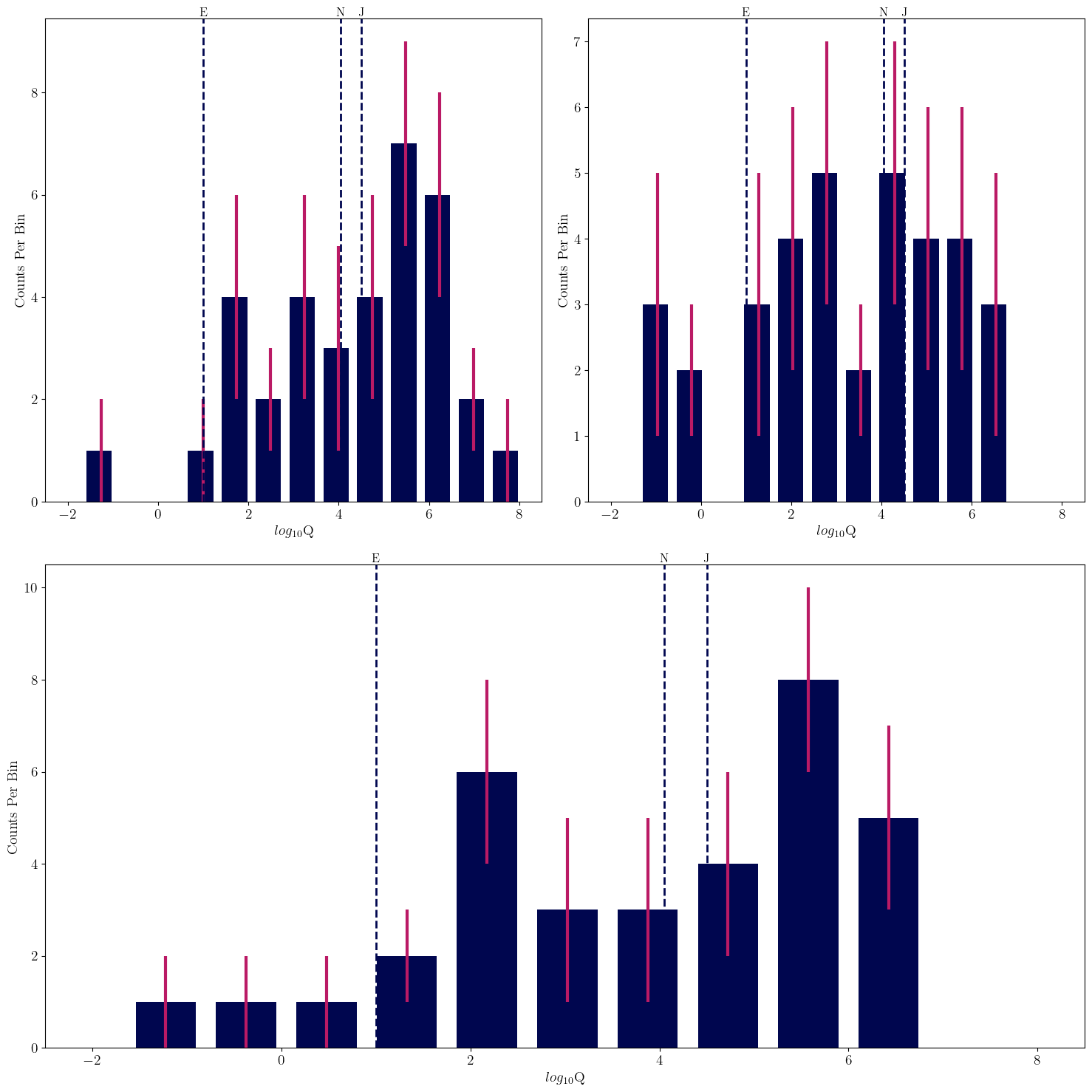}
    \caption{\textbf{\textit{Top:} Inferred $Q$ values when inner (left)/outer (right) planets are assumed to drive resonant repulsion.} Dashed lines mark $Q$ values estimated for Earth, Neptune, and Jupiter. The solid magenta vertical lines for each bin correspond to the $1-\sigma$ bounds on the count of each bin observed in $N=5000$ bootstrap trials. 
    \textbf{\textit{Bottom:} The distribution of inferred $Q$ values within near-resonant short-period planets when randomly selecting one planet from each pair to contribute.} The $1-\sigma$ confidence intervals are as in the top panels.}
   \label{fig:Q_bootstrap}
\end{figure*}

We evaluate the above expression to estimate a prospective $Q$ for each inner member of our planetary pairs. We employ bootstrap resampling with replacement to assess uncertainty and implement $N=5000$ trials. Each trial generates a histogram of estimates for $Q_1$, and we maintain constant bin boundaries over the trials. The 15th and 85th percentiles of the counts for each bin are adopted as approximate to the one-sigma uncertainties for the histogram counts in each bin, and these limits are plotted as solid bars in Figure \ref{fig:Q_bootstrap}. The top left panel of Figure \ref{fig:Q_bootstrap} presents the resulting distribution of $Q_1$ values.

The foregoing process is then repeated to estimate the distribution of values, $Q_2$, which result if the outer member of each pair is assumed to be dissipating via obliquity tides at a rate sufficient to generate the entire observed repulsion, $\Delta_{mig}$, over the system's age. The results are plotted in the top right panel of Figure \ref{fig:Q_bootstrap}.

\begin{equation}\label{eq:outer_Q}
   Q_2   = \frac{9\mu_1 t}{4} \Delta_{mig}^{-3}\left[(2+\beta)(C_{e2} X_2f_2^2) - \frac{1}{2} C_{a2} X_2f_2^2\right].
\end{equation}

Given the estimated distributions of $Q$ stemming from the two different assumptions (inner versus outer planet undergoing dissipation), we next evaluate a scenario where one planet in each pair is randomly selected to be the dissipating member. We use the above-described bootstrap procedure to generate the bottom panel of Figure \ref{fig:Q_bootstrap}. 

The known presence of two distinct populations in the planetary radius distribution motivates us to assess whether there is also evidence for a bimodal structure in \ref{fig:Q_bootstrap}. We employ Hartigan's non-parametric Dip test \citep{Hartigan} to assess whether our distribution of estimated $Q$ values diverges from unimodality. The test measures the greatest discrepancy between the empirical distribution function of the data and the unimodal distribution function that most closely approximates the data. The ``dip statistic'' indicates the degree of divergence from unimodality, with a larger dip providing more evidence of multimodality. The results are shown in Table \ref{tab:stats}. The p-values in all three cases are greater than 0.05, indicating that based on this test, we cannot reject the null hypothesis that the samples are all drawn from a single distribution.

If a bimodal signal exists in the current data, its presence is obscured by both the uncertainty in the planetary parameters used in the analysis and the assumption that in each case only one planet is contributing. Because there is a large dynamic range in $Q$ and a narrow separation between the expected dissipation regimes, it will take a larger sample to reveal a statistically significant signal.

\begin{table}
    \centering
    \caption{Dip Statistic Calculations}
    \begin{tabular}{l|c|c}
      Sample   & Dip Statistic & p-value \\
      \hline
      Inner Planets   &  0.04 & 0.91\\
      Outer Planets & 0.06 & 0.32\\
      Combined & 0.03 &  0.90\\
     \end{tabular}
\label{tab:stats}
\end{table}

\section{Summary and Discussion}\label{sec:discussion}

In this paper, we explored the regimes of tidal quality factors of short-period exoplanets under the hypothesis that obliquity tides are a driving source of dissipation in the planets near the 2:1 and 3:2 integer commensurabilities. \cite{2019NatAs...3..424M} demonstrated that compact multi-planet systems are frequently susceptible to secular-spin orbit resonances, which lead to obliquity excitation and tidal dissipation. The dissipation occurs at different rates for different planetary compositions. If the two hypothesized populations of rocky and gaseous planets exist, aggregation of estimates of their tidal $Q$'s should eventually reveal two distinct groups. 

We estimated individual $Q$ values using a relation between the dissipation strength and the distance from perfect period-ratio commensurability under assumptions that either the inner planets are in the high-obliquity state or the outer planets are in the high-obliquity state. We performed random pairwise selection to create a cumulative distribution. We argue that this procedure can potentially provide a robust dynamical tool to probe the dissipation regimes that characterize the short-period planets in multiple-transiting systems.

While there is as-yet no signal indicating bimodality in the distribution, we predict that as more near-commensurate multi-transiting multi-planet systems are identified, two peaks at $Q\approx 10^3$ and $10^4$ would emerge if geophysically distinct populations of super-Earth ($Q=10^3$) and sub-Neptune ($Q=10^4$) planets \citep{Murray1998} exist.

Our study assumes that the relevant orbital and physical parameters of the systems are known well enough to estimate the tidal $Q$ values of the planets. The parameters of these systems, however, are not all precisely constrained; in particular, the planetary masses have significant uncertainties. To establish how many planets would be needed to place the existence of two peaks on firm statistical ground, we created a bimodal distribution centered on the locations of the two peaks with a $\sigma$ of 0.5. We increased the sample size until the p-value of the dip test was less than 0.05. The necessary sample size is around 800 planets. Given the ongoing productivity of TESS and the pending launch of PLATO, which is expected to discover 7000 planets
\footnote{\url{https://sci.esa.int/documents/33240/36096/1567260308850-PLATO_Definition_Study_Report_1_2.pdf}}, it is reasonable to expect that the population will reach the required size within the next five to ten years.   

\section*{Acknowledgments}
E.L. thanks Garrett Levine, Tiger Lu, Konstantin Gerbig, Sam Cabot, Isabel Medlock, and Harrison Souchereau for helpful and thought-provoking conversations.

\facility {Exoplanet Archive, ADS}

\software{\texttt{numpy} \citep{harris2020array};
\texttt{scipy} \citep{2020SciPy-NMeth};
\texttt{pandas} \citep{reback2020pandas}
}

\bibliography{main}{}
\bibliographystyle{aasjournal}

\end{document}